\documentclass[aps,prb,twocolumn,groupedaddress,showpacs]{revtex4}

\usepackage{graphicx}

\begin{document}


\title{The Generic, Incommensurate Transition in the two-dimensional Boson Hubbard Model}


\author{Fabien Alet$^{(1,2)}$}
\email{alet@phys.ethz.ch}

\author{Erik~S.~S\o rensen$(3)$}
\email{sorensen@mcmaster.ca}
\homepage{http://ingwin.physics.mcmaster.ca/~sorensen}

\affiliation{$^{(1)}$Theoretische Physik, ETH Z\"urich, CH-8093 Z\"urich, Switzerland}
\affiliation{$^{(2)}$Computational Laboratory, ETH Z\"urich, CH-8092 Z\"urich,
  Switzerland}
\affiliation{$^{(3)}$Department of Physics and Astronomy, McMaster University, Hamilton, ON, L8S 4M1 Canada}
\date{\today}

\begin{abstract}
The generic transition in the boson Hubbard model, occurring at an incommensurate chemical potential, is
studied in the link-current representation using the recently developed directed geometrical worm algorithm.
We find clear evidence for a multi-peak structure in the energy distribution for finite lattices,
usually indicative of a first order phase transition. However, this multi-peak structure is shown to 
disappear in the thermodynamic limit revealing that the {\it true} phase transition is second order.
These findings cast doubts over the conclusion drawn in a number of previous works considering
the relevance of disorder at this transition.
\end{abstract}


\maketitle



%



\section{Introduction}
Quantum phase transitions occurring in bosonic systems have experienced a surge of
interest lately, due to recent experiments showing clear evidence for a quantum phase
transition occurring in an optical lattice~\cite{greiner}. In these experiments, one-,
two- and three-dimensional lattice structures are imposed on the bosonic system by using lasers to
create standing wave patterns. The lattice parameters and the lattice structure of this artificial
lattice can therefore be experimentally tuned, creating an almost ideal setting for studying quantum
phase transitions occurring in bosonic systems. Here we shall mainly be concerned with the case of
a two-dimensional system. In the simplest setting the quantum phase transition is in this case
believed to occur between a Mott insulating (MI) phase where the number of bosons per site
is fixed and the phase therefore incompressible and a superfluid (SF) phase where the bosons
freely hop throughout the lattice. However, several other phases are also possible~\cite{opttheo}.

In the experiments described above the constituent atoms are clearly bosons and quantum phase transitions observed
in $^4$He films~\cite{crowell} should be in the same superfluid-insulator (SF-I) universality class.
A detailed scaling theory of the SF-I phase transition arising in two-dimensions as described by the
bosonic version of the Hubbard model was developed in a seminal
paper by Fisher et. al~\cite{Fisher89c}, in particular the transition in the presence of disorder
was investigated and it was pointed out that for the SF-I transition in a disordered systems the insulating phase 
would in almost all situations not be a Mott insulator but instead a compressible Bose glass.
Subsequent work showed that the same physics should apply also to quantum phase transitions occurring
in superconducting systems dominated by a diverging phase coherence length~\cite{Fisher90a,Fisher90b} making the
fermionic degrees of freedom irrelevant at the critical point. The reasoning being that, at the scale
of the diverging phase coherence length the size of the individual Cooper pairs would be negligible and the underlying
physics should therefore be dominated by the bosonic degrees of freedom. The scaling theory for the SF-I
transition~\cite{Fisher89c}, mainly focusing on phase fluctuations, should therefore also apply to
the two-dimensional superconductor-insulator (SC-I) transition. Ensuing experiments showed support
for this scaling picture arising in superconducting films~\cite{Liu,Markovic,goldman} and several similar systems
such as Josephson Junction arrays~\cite{Zant}. This scaling theory predicts that in two dimensions the
quantum phase transitions for both the SF-I as well as the SC-I transition should take place directly from
the superfluid or superconducting phase into the insulating phase with no intervening metallic phase.
However, more recent experiments~\cite{Kapitulnik} and theoretical studies~\cite{Wagenblast,Dalidovich,Das,Phillips} have 
pointed to the possibility of a metallic phase occurring in two-dimensional bosonic systems in particular in the presence
of dissipation~\cite{Wagenblast,Dalidovich}.

Only at very special points does the SF-I occur at a commensurate filling factor. At these points the
scaling theory of Fisher {\it et al.}~\cite{Fisher89c} predicts that the transition is in the $d+1$ dimensional
$XY$ class dominated by phase-fluctuations. More often the transition occurs at an incommensurate chemical
potential which is in a {\it different} universality class. This more {\it generic} transition is expected
to be mean-field like~\cite{Fisher89c} and is the focus of the present paper.

Initial quantum Monte Carlo (QMC) simulations~\cite{Krauth,Trivedi,Zhang} performed directly on the boson Hubbard model showed
clear evidence for a direct SF-I transition. These simulations were constrained to systems with a fixed particle number
and therefore always incompressible. Implicitly only the commensurate transition was studied. Subsequent studies~\cite{Sorensen,Wallin} exploiting a mapping to the link-current model
removing this constraint focused on the phase transition in the presence of strong disorder and showed clear evidence for
a transition from the superfluid to a {\it Bose glass} phase, as did studies
at fixed densities~\cite{Trivedi}.

A point of controversy has been the phase transition occurring at {\it weak} disorder. 
A generalization~\cite{Chayes}
of the Harris criterion shows that the transition is stable towards disorder if $\nu\ge 2/d$. In $d=2$
the $d+1$ dimensional $XY$ does presumably not satisfy this inequality and neither does
the mean-field $\nu=1/2$ predicted to occur at the {\it generic} transition. 
The scaling theory~\cite{Fisher89c} therefore predicts that in all cases  the Bose-glass phase intervenes
between the superfluid and insulating phases. This prediction was however contradicted by numerical
simulations showing evidence for a direct SF-I transition at a commensurate chemical potential in 
the presence of weak disorder~\cite{Kisker}. Furthermore, Lee {\it et al.}~\cite{Park,Lee,Leerr} reported evidence
for a direct SF-I transition also at incommensurate chemical potentials. From these studies one would
conclude that the SF-I transition should be {\it stable} towards the introduction of disorder,
contradicting the scaling theory. 

The recent development of worm algorithms~\cite{Prokworm,bosons} have allowed for the study of significantly
larger system sizes and Prokof'ev and Svistunov~\cite{ProkDis1,ProkDis2} have shown evidence for the relevance
of disorder at the transition occurring at commensurate chemical potentials with the universal behavior
only setting in at very large system sizes. The existence of a multicritical line proposed in Ref.~\onlinecite{Lee,Leerr}
was also ruled out by these studies. However, the controversy surrounding the incommensurate case still remains.
In fact, to our knowledge, no strong numerical evidence exists supporting the evidence of mean-field like
exponents even in the {\it absence} of disorder for the generic transition, the results implicit in previous
studies of the transition in the presence of disorder~\cite{Park,Lee,Leerr} as well as studies considering
longer range interactions and the possibility of supersolid phases~\cite{vanOtterlo94,vanOtterlo95} being limited to restrictively small
lattice sizes.  The results including longer range interactions were subsequently criticized~\cite{Batrouni95}.
All these studies~\cite{Kisker,vanOtterlo94,vanOtterlo95,Park,Lee,Leerr,ProkDis1,ProkDis2} were performed using the
link-current representation which we shall also use in this study.

In light of this wealth of experimental and theoretical studies it is perhaps surprising that
fundamental aspects of the SF-I transition as it occurs at an incommensurate chemical potential are still 
controversial. This so called {\it generic} 
transition is the one most likely to describe real experiments and in the
present paper we present large-scale numerical results using a recently
developed geometrical worm algorithm, capable of yielding precise results for
lattice sizes largely surpassing previous studies, thereby allowing us to shed new
light on this transition. In particular we show that, due to the fact that this
transition is dominated by fluctuations of the particle number, simulations on
finite lattice will in most cases show clearly defined peaks in the energy histograms reminiscent of a
first-order transition. Only in the thermodynamic
limit do the peaks coalesce and a second-order transition is recovered. As stressed by Prokof'ev and
Svistunov~\cite{ProkDis1,ProkDis2} for the commensurate transition, extreme care should therefore be taken when
applying a finite-size scaling analysis. For the relative small lattice sizes
used in the studies by Lee {\it et al.}~\cite{Lee,Leerr} the true influence of
disorder is therefore likely even further obscured by these finite-size energy
gaps associated with the first-order transition.

In the remainder of this section we discuss the Boson Hubbard model and the particular link-current representation
that we use for this study as well as the associated phase-diagram. Section~\ref{sec:numerical} describes
the numerical techniques employed and Section~\ref{sec:results} focuses on our
results showing features characteristic of a first-order transition for finite
lattices, which are found to disappear in the thermodynamic limit. 
We conclude with a discussion in Section~\ref{sec:conclusion}.

\subsection{The model}
The simplest model we can write down for interacting bosons
must at least include an on-site repulsion term $U$ as well as a competing hopping
term parametrized by the hopping strength $t$. In the absence of the on-site repulsion term a
bosonic system would always condense and would always be superfluid at $T=0$.
Since we in the present paper in particular focus on the generic transition occurring at an incommensurate
filling we also include a chemical potential, $\mu$. We thus arrive at the
well known boson Hubbard model:
\begin{equation}
H_{\rm bH}=\sum_{\bf r}\left(\frac{U}{2}\hat n_{\bf r}(\hat n_{\bf r}-1) -
\mu\hat n_{\bf r}\right)- \frac{t}{2}\sum_{\langle {\bf r},{\bf r'}\rangle }
(\hat \Phi^\dagger_{\bf r}\hat \Phi_{\bf r'}+c.c) \ .
\label{eq:HbH}
\end{equation}
Here $\Phi^\dagger_{\bf r}$ ($\Phi_{\bf r}$) is the creation (annihilation)
operator at site ${\bf r}$ and 
$\hat n_{\bf r}=\hat \Phi^\dagger_{\bf r}\hat \Phi_{\bf r}$ is the number operator.
At $t=0$ the bosons are completely localized in Mott insulating phases while it can be shown~\cite{Fisher89c}
that a non-zero $t$ eventually gives rise to a superfluid phase with the Mott insulating phase persisting in 
a series of lobes into the superfluid. In the MI phase the particle number is fixed and only at the tip of the
MI lobes does the density not change at the SF-I transition. This transition is therefore dominated by
phase-fluctuations and is in the $d+1$-dimensional $XY$ class. The generic transition, occurring at incommensurate
filling factors, is however dominated by fluctuations in the particle number and is expected to be characterized
by mean-field exponents~\cite{Fisher89c}.

We can simplify the Hamiltonian Eq.~(\ref{eq:HbH}) by integrating out amplitude fluctuations.
We first set $\hat \Phi_{\bf r}\equiv|\hat \Phi_{\bf r}|e^{i\hat \theta_{\bf r}}$,
where $\theta_{\bf r}$ is the phase of a quantum rotor.
By performing the integration~\cite{Sorensen,Wallin}
$H_{\rm bH}$ then becomes equivalent to the (N=2) model of quantum rotors $(\cos(\theta_{\bf r}),\sin(\theta_{\bf r}))$,
that describes a wide range of phase transitions
dominated by phase-fluctuations:
\begin{equation}
H_{\text{qr}}=
\frac{U}{2}\sum_{\bf r}
\left( \frac{1}{i}\frac{\partial}{\partial
\theta_{\bf r}} \right)^2
+i\sum_{\bf r} \mu
\frac{\partial}{\partial \theta_{\bf r}}
-t\sum_{\langle {\bf r},{\bf r'}\rangle }
\cos(\theta_{\bf r}-\theta_{\bf r'}).
\label{eq:hqr}
\end{equation}
Here, $t$ is the renormalized
hopping strength and $\frac{1}{i}\frac{\partial}{\partial\theta_{\bf r}} =L_{\bf r}$ is the angular momentum
of the quantum rotor. The angular momentum can be thought of as describing the deviation of the particle
number from its mean, $L_{\bf r}\simeq n_{\bf r}-n_0$.
Hence, an equivalent
Josephson junction array form of this Hamiltonian is:
\begin{eqnarray}
H_{\text{JJ}}&=&
\frac{U}{2}\sum_{\bf r}
\left(n_{\bf r}-n_0 \right)^2
-i\sum_{\bf r} \mu ( n_{\bf r}-n_0)\nonumber\\
&&-t\sum_{\langle {\bf r},{\bf r'}\rangle }
\cos(\theta_{\bf r}-\theta_{\bf r'}).
\label{eq:hJJ}
\end{eqnarray}
If $H_{qr}$ is written in its Villain~\cite{Villain} form we obtain
a classical model~\cite{JKKN,FisherLee,Sorensen,Wallin} in the same universality class
where the Hamiltonian is written in terms of integer currents defined on the links of
a lattice, ${\bf J}=(J^x,J^y,J^\tau)$ that we shall use in this study. One finds~\cite{Sorensen,Wallin}:
\begin{equation}
H=\frac{1}{K} {\sum_{({\bf r},\tau)}}
\left[\frac{1}{2} {\bf J}_{({\bf r},\tau)}^{2}- \mu J_{({\bf r},\tau)}^\tau\right].
\label{eq:hV}
\end{equation}
In this (2+1) dimensional classical Hamiltonian
the link-current variables describe the {\it total}
``relativistic" bosonic current which has to be conserved on the space-time lattice and therefore has
to be divergenceless, ${\bf \nabla \cdot J} = 0$.
The link-current variables take on integer values $J^x,J^y,J^\tau=0,\pm 1, \pm 2, \pm 3 \ldots$. Intuitively
$J^x_{({\bf r},\tau)}$, $J^y_{({\bf r},\tau)}$ describe the integer number of bosons hopping in the $x$ or $y$ direction
from the site ${\bf r}$ at imaginary time $\tau$ where as $J^\tau_{({\bf r},\tau)}$ denote the number of bosons
that remain at the site ${\bf r}$ at imaginary time $\tau$. $K$ is the
effective temperature, varying like $t/U$ in the quantum rotor model.

In this representation, when $K=0$, there is an integer number $n_0=\langle J^\tau \rangle$ of bosons on each
site in order to minimize the energy per site. All link variables of the
classical 3D model vanish in the space directions. The ``ground state'' is
composed of $n_0$ bosons per site when the chemical potential is in the
interval $n_0-1/2 < \mu < n_0+1/2$, and the compressibility $\kappa$ defined as
$\kappa=\frac{\partial n}{\partial \mu}$ is zero. This is the incompressible Mott Insulating phase with precisely
$n_0$ bosons per site.

In the other limit $K\rightarrow \infty$, the bosons are free to hop on the
lattice and condensate in the lowest energy mode $k=0$. We have
off-diagonal long range order and the system is in a superfluid phase. The
compressibility $\kappa$ is non-zero since the boson number is fluctuating.

\begin{figure}
\includegraphics[clip,width=8cm]{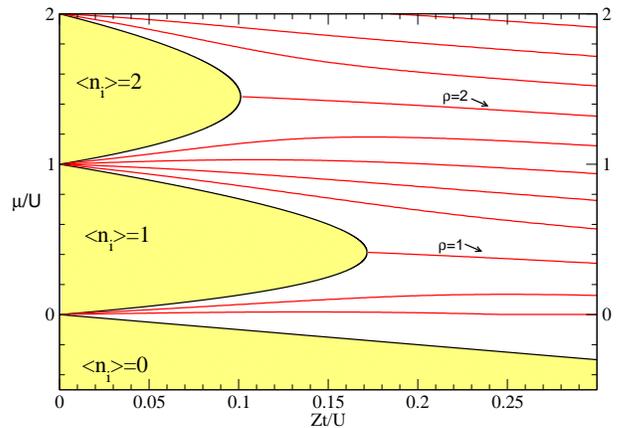}
\caption{Mean field phase diagram in the $(\mu/U,Zt/U)$ plane for the boson Hubbard model
(\ref{eq:HbH}). }
\label{fig:Tphase}
\end{figure}

\subsection{Phase Diagram for the pure case}
The phase-diagram of Eq.~(\ref{eq:HbH}) has been obtained using several equivalent approaches.
Sheshadri {\it et al.}~\cite{Sheshadri} decouple the hopping term in the following manner:
\begin{equation}
\hat \Phi^\dagger_{\bf r}\hat \Phi_{\bf r'}=\langle\hat \Phi^\dagger_{\bf r}\rangle\hat \Phi_{\bf r'}
+\langle\hat \Phi_{\bf r'}\rangle\hat \Phi^\dagger_{\bf r}-\langle\hat \Phi^\dagger_{\bf r}\rangle\langle\hat \Phi_{\bf r'}\rangle.
\end{equation}
Using this decoupling and writing $\psi^\star=\langle\hat \Phi^\dagger_{\bf r}\rangle, \psi=\langle\hat \Phi_{\bf r}\rangle$ we see
that Eq.~(\ref{eq:HbH}) can be written as follows:
\begin{eqnarray}
H_{\rm bH}&\simeq&\sum_{\bf r}H^{MF}_{\bf r}\nonumber\\
H^{MF}_{\bf r}&=&\left(\frac{U}{2}\hat n_{\bf r}(\hat n_{\bf r}-1) -\mu\hat n_{\bf r}\right)\nonumber\\
& &- Zt(\psi^\star\hat \Phi_{\bf r}+\psi\hat \Phi^\dagger_{\bf r}-|\psi|^2)\ .
\label{eq:Hmf}
\end{eqnarray}
Here $Z$ is the coordination number.
In the number basis $H^{MF}_{\bf r}$ is non-diagonal but straightforward to diagonalize
out to fairly large occupation numbers. In this truncated basis the ground-state energy
can be determined and the optimal value of $\psi$ determined where the energy is minimized. 
This procedure explicitly yields the ground-state wave-function at mean-field level and hence directly
the density. The superfluid density can be determined from $\rho_s=|\psi|^2$. This approach is equivalent
to preceding work by Fisher {\it et al.}~\cite{Fisher89c} who showed that the Mott insulating lobes with occupation $n$
could be determined by considering an action of the typical Landau form $S_\infty(\psi)=\beta N\frac{1}{2}r|\psi|^2+\ldots$,
with $r$ given explicitly at $T=0$ by
\begin{equation}
r=2\frac{U}{Zt}-2\left(\frac{n+1}{n-\mu/U}+\frac{n}{-(n-1)+\mu/U}\right)\ .
\end{equation}
The transition from the Mott insulator with occupation $n>0$ to the superfluid is in this mean field approach given by $r=0$ with solution~\cite{Freericks}:
\begin{equation}
\frac{\mu}{U}=n-\frac{1}{2}-\frac{x}{2}\pm \frac{1}{2}\sqrt{1-2x(2n+1)+x^2}\ .
\label{eq:mlobes}
\end{equation}
Here, $x=Zt/U$. For the Mott insulator with $n=0$ the transition is simply given by $\mu/U=-Zt/U$.
Note that in Eq.~(\ref{eq:HbH}) the coefficient in front of $-\sum_i n_i$ is $\mu/U+1/2$. Hence the Mott insulating lobes
are not centered around their corresponding chemical potential, but are off set by 1/2.
The resulting phase diagram is shown in Fig.~\ref{fig:Tphase} and is in quite good agreement with
detailed strong-coupling expansions~\cite{Freericks,Niemeyer}. In Fig.~\ref{fig:Tphase} Eq.~(\ref{eq:mlobes}) are
shown along with detailed calculations using Eq.~(\ref{eq:Hmf}) using the approach of Ref.~\onlinecite{Sheshadri}.

It is now straightforward to consider the Mott insulating lobes present in the quantum rotor model, Eq.~(\ref{eq:hqr}), Eq.~(\ref{eq:hJJ}).
This can be done simply by studying the limit $n\to\infty$ of Eq.~(\ref{eq:mlobes}) considering that
$nZt/U$ remains finite. One immediately finds that the lobes are given by:
\begin{equation}
\frac{\mu}{U}=\frac{1}{2}\pm \frac{1}{2}\sqrt{1-4\frac{nZt}{U}}\ .
\end{equation}

\begin{figure}
\includegraphics[clip,width=8cm]{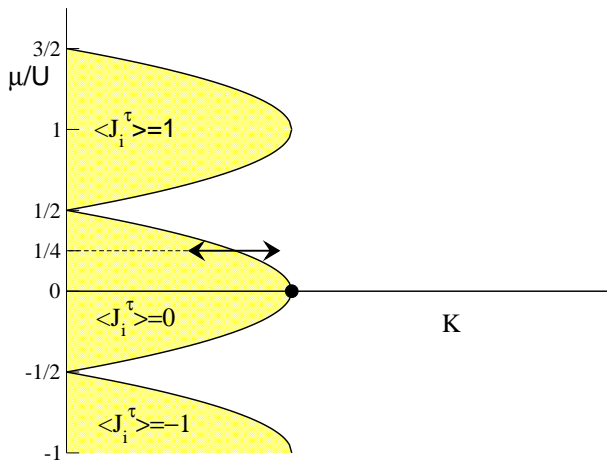}
\caption{Schematic phase diagram in the $(\mu/U,K)$ plane for the link-current
Hamiltonian (\ref{eq:hV}). We have simulated the transition between the Mott
insulator and the SuperFluid phase at $\mu=1/4$ (arrow in the figure).}
\label{fig:Kphase}
\end{figure}

For the link-current model Eq.~(\ref{eq:hJJ}) the only non-trivial interaction comes through 
the global divergence-less constraint and a mean-field treatment is less straight forward.
However, as outlined in Ref.~\onlinecite{Wallin} the coupling in the link-current model Eq.~(\ref{eq:hJJ})
is related to $t/U$ of the quantum rotor model Eq.~(\ref{eq:hqr}) and the overall shape of the
phase-diagram must therefore be the same. Since the particle number now describes the deviation
from the mean the resulting phase-diagram therefore becomes completely periodic in $\mu/U$ and 
is schematically indicated in Fig.~\ref{fig:Kphase}.

In a previous study~\cite{bosons}, we studied the quantum phase transition
at the tip of the lobe $\mu=0$ (black dot in figure (\ref{fig:Kphase})), where
it is known that this model is in the universality class of the 3D XY
model~\cite{Fisher89c,Cha}. 
We gave a very precise estimate of the critical point $K_c=0.33305(5)$ and found a value
$\nu=0.670(3)$ for the correlation length critical exponent, in perfect
agreement with the 3D XY universality class~\cite{Campostrini01}. The
dynamical critical exponent $z$ is equal to unity in this case.

In this work, we will focus on the pure case where the chemical potential is
the same for all sites $\mu_{\bf r}=\mu$. In particular, we will present
results obtained for a value $\mu=1/4$ of the chemical potential indicated by
the arrow in Fig~(\ref{fig:Kphase}).
At this point, the quantum phase transition is expected
from scaling theory~\cite{Fisher89c} to have a dynamical
critical exponent $z=2$, and mean field values for other exponents (in
particular $\nu=1/2$). As we shall see, our simulations confirm this with a good
accuracy in the thermodynamic limit, but we find finite size effects that are
strongly reminiscent of a first-order phase transition.

This has important implications for studies considering the relevance of disorder
at this generic transition~\cite{Park,Lee,Leerr} since our results show that for
the lattice sizes used in these studies the phase-transition appears first-order like.
 
\section{Numerical method and simulations~\label{sec:numerical}}

\subsection{Numerical method}

We perform Quantum Monte Carlo simulations of the model~(\ref{eq:hV}) with the
recently proposed geometrical worm algorithm~\cite{bosons,directed} in its "directed" version~\cite{directed}.
We refer the interested reader to these references for more detailed
explanations on this numerical technique. We also note that our algorithms are
closely related to the classical worm algorithm by Prokof'ev and
Svistunov~\cite{Prokworm} (a comparison between the two approaches has been
made in Ref.~\cite{directed}).
The most important point to
underline is that these non local Monte Carlo algorithms permit the study of much
larger systems with much
higher precision than what was previously possible using
local update schemes, thereby getting closer to the thermodynamic limit and
allowing for much more precise estimates of the critical properties of the model.

\subsection{Simulations}
\label{sec:sim}

We are considering a quantum phase transition with two different
correlation lengths: $\xi$ is the correlation length in spatial
directions ($x,y$) and  $\xi_\tau$, the correlation length in the $\tau$
(imaginary-time) direction. Usually one defines $\xi_\tau\sim\xi^z$, thereby defining the
dynamical critical exponent $z$, not necessarily equal to unity.
To respect this anisotropy between space and time directions, 
it is necessary to simulate the
the model~(\ref{eq:hV}) on
lattices of sizes $L$x$L$x$L_\tau$. 
Periodic boundary conditions
are here assumed and the length of the lattice in the $\tau$- direction has to  be chosen
such that $L_\tau=aL^z$, with $a$ being the aspect ratio constant throughout the simulations. 
This follows from the fact that any finite-size scaling function will
be a function of {\it two} arguments:
\begin{equation}
f(\xi/L,\xi_\tau/L_\tau)=g(\xi/L,a).
\end{equation}
It is therefore necessary to keep the aspect ratio $a$ constant in order to observe scaling.
The value of the
dynamical critical exponent is {\it a priori} unknown, and one usually has to  try
several values for $z$ to check the validity of theoretical predictions.

Most of the data for different values of the effective temperature $K$
were obtained with reweighting techniques~\cite{Ferrenberg88} by large runs
(of the order of $5\times 10^7$ worms) at a single value of $K$ although we in some
cases combined simulations at several values of $K$ using multi-histogram techniques. We checked
every time that our data were in the range of validity of the reweighting. The
error bars are obtained with standard jackknife resampling
techniques~\cite{Efron}.

Among several possible thermodynamic variables, we have calculated two quantities 
capable of distinguishing the
different phases of the system~\cite{Cha,Sorensen}. The first quantity is the
stiffness $\rho$ which characterizes the response of the system to a twist in
the boundary condition in the real space direction. In
terms of the link variables, the stiffness is calculated as~\cite{Cha,Sorensen,Wallin} :
\begin{equation}
\rho = \frac{1}{L_\tau} \langle n_x^2 \rangle
\end{equation}
where $n_x=\frac{1}{L}\sum_{{\bf r},\tau} J_{({\bf r},\tau)}^x$ is the
winding number in the $x$ direction.

The second quantity is the compressibility $\kappa$, characterizing 
the fluctuation of the number of bosons in the system. This
quantity is written as~\cite{Sorensen,Wallin}:
\begin{equation}
\kappa = \frac{L_\tau}{L^2} [ \langle n_\tau^2 \rangle - \langle n_\tau \rangle^2]
\end{equation}
with $n_\tau=\frac{1}{L_\tau}\sum_{{\bf r},\tau} J_{({\bf r},\tau)}^\tau$
being the winding number in the time direction. This last quantity can be
interpreted as the ``number of bosons'' in the systems. Please note that
$\langle n^\tau \rangle$ is in general non-zero when we consider a non-zero chemical
potential.

We will use finite-size scaling relation for these two quantities to
localize critical points and characterize them. In two dimensions and near a
critical point $K_c$, the scaling theory~\cite{Fisher89c,Sorensen,Wallin} predicts the
following finite size scaling forms for the stiffness and the compressibility :
\begin{equation}
\rho=L^{2-d-z} \tilde{\rho}(L^{1/\nu}\delta,a),
\label{eq:rhoscal}
\end{equation}
\begin{equation}
\kappa=L^{z-d} \tilde{\kappa}(L^{1/\nu}\delta,a),
\end{equation}
where $\delta=|K-K_c|$ is the distance to the critical point, $\nu$ the
correlation length critical exponent and where $\tilde{\rho}$ and
$\tilde{\kappa}$ are scaling functions.

From these scaling forms, we see that right at the critical point $K_c$, in two
dimensions $d=2$, the
quantities $\rho L^z$ and $\kappa L^{2-z}$ must be independent of the system
size if we choose the aspect ratio $a$ to be constant for different lattice
sizes. Thus a plot of $\rho L^z$ or $\kappa L^{2-z}$ versus $K$ for different
system sizes should show a crossing of the different curves at a single
transition temperature $K_c$. 

Moreover, by simply differentiating the equation (\ref{eq:rhoscal}) with
respect to the coupling $K$, we easily see that at the critical point $K_c$,
we have (keeping the aspect ratio constant)
\begin{equation}
\frac{d\rho}{dK} L^z \sim L^{1/\nu}
\label{eq:drho}
\end{equation}
which is used to get the correlation length exponent. We find this
way of determining $\nu$ much preferable to the traditional data 
collapse~\cite{Kisker,Cha,vanOtterlo94,vanOtterlo95,Park,Lee,Leerr}, 
which leads to significantly more uncertainty in the determination of
the critical exponents.

The derivative of $\rho$ with respect to the effective temperature $K$ can
be obtained by a numerical derivation of the curve $\rho(K)$ or more
preferably by calculating the thermodynamic derivative during the Monte
Carlo simulations :
\begin{equation}
\frac{d\rho}{dK}=\langle \rho E \rangle - \langle \rho \rangle 
\langle E \rangle 
\end{equation}
where $E$ is the total energy.

\section{Results~\label{sec:results}}

Throughout the rest of this work, we will show results obtained with the ``directed" geometric worm
algorithm~\cite{bosons,directed} for the link-current model~(\ref{eq:hV}) at an incommensurate
chemical potential, $\mu=1/4$. We exclusively consider the two dimensional case with {\it no} disorder.

\subsection{Determination of critical point}

First, we address the question of the value of the critical point $K_c$ and
of critical exponents. The scaling theory~\cite{Fisher89c} predicts
$z=2$ for the dynamical critical exponent. A plot of $\rho L^2$ or $\kappa$
should then show a crossing of the curves for different systems sizes at a
single point. We have calculated this quantity near the previous
estimate~\cite{vanOtterlo94,vanOtterlo95} of the transition point $K_c\simeq 0.283(3)$ for
different lattice sizes for different values of the aspect ratio~:~$a=1/32,
1/16, 1/8, 1/4, 1/2, 1$. The small aspect
ratios allows us to treat larger systems in the real space directions,
gaining in precision. For all values of $a$, we simulated large
systems, up to more than $3\times 10^6$ lattice variables. The maximum size used for
different aspect ratios is indicated in table~\ref{tab.1}. A more thorough discussion
of the influence of the aspect ratio follows
in section~\ref{sec:aspect}. Previous work~\cite{vanOtterlo94,vanOtterlo95} using a local
algorithm was limited to a maximum size $L=10$ with an aspect ratio $a=1/4$
(that is to say lattices of size $10$x$10$x$25$), whereas with the help of
the geometrical worm algorithm we have been able to simulate lattices of size up to
$88$x$88$x$242$ for the smallest aspect ratio used $a=1/32$ and in principle even
larger lattice sizes could be studied. However, as we shall see, features present
in the results obtained using these lattice sizes already tell us that the extrapolation
to the thermodynamic limit will be difficult.

\begin{figure*}
\includegraphics[clip,width=16cm]{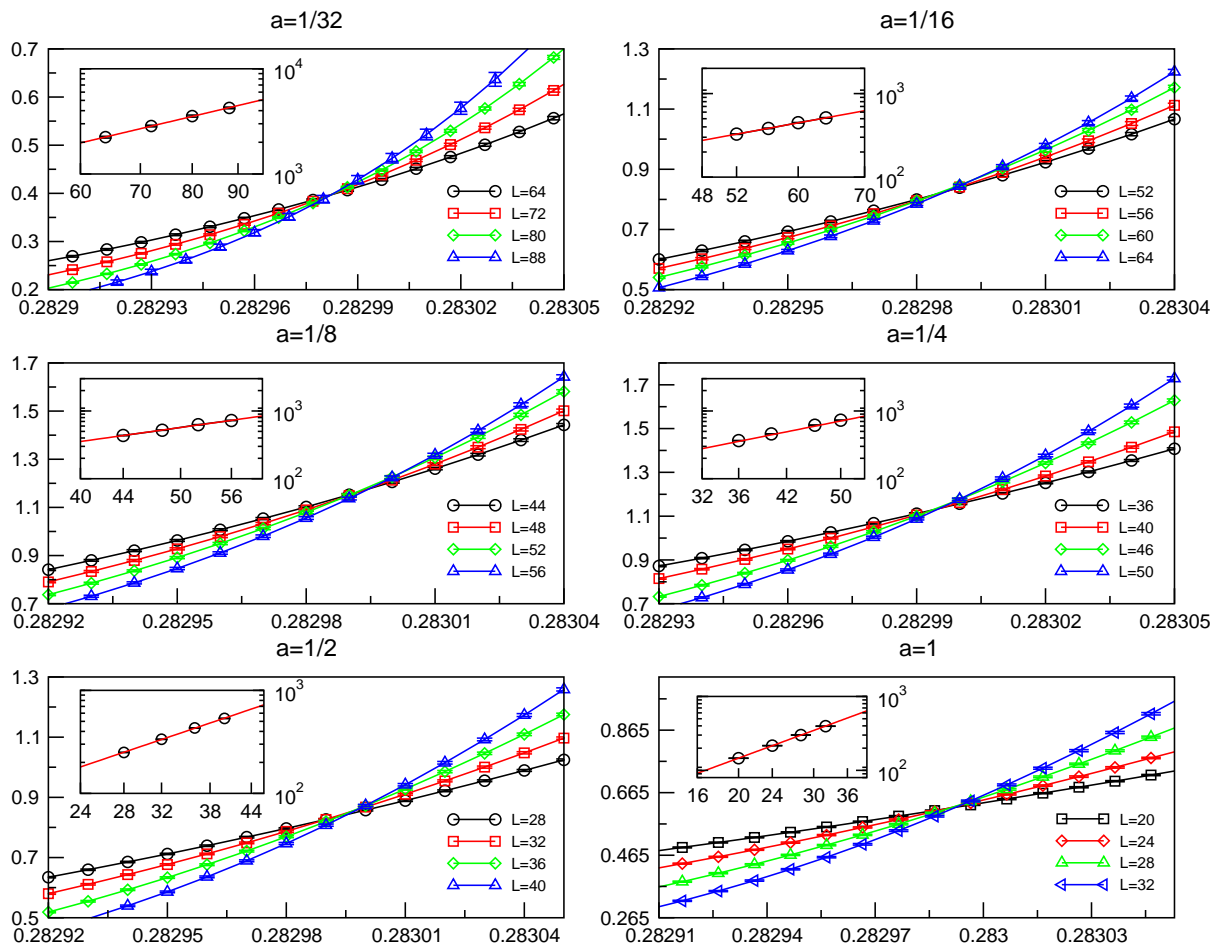}
\caption{The scaled stiffness $\rho L^2$ (vertical axis) versus coupling constant $K$
  (horizontal axis) for different
  values of aspect ratios $a$ and system sizes $L$. Lines are guides to the
  eye. In all cases, the curves
  for different systems sizes cross at the critical point, giving a precise
  estimate of $K_c$. Insets: $L^2\frac{d \rho}{dK}$ at
  $K_c$ as a function of system size $L$ in log-log scale. Solid lines are power-law fits.}
\label{fig:RhoL2}
\end{figure*}

In Figure~\ref{fig:RhoL2} we show results for the dependence of $\rho L^2$ on $K$ for
different values of $L$ and the set of different aspect ratios. We see a very good
crossing in all cases, and the values 
of $K_c$ (listed in table~\ref{tab.1}) only show a very slight variation
with the aspect ratio, and all converge to give an estimate of $K_c=0.28299(2)$. As one would
expect, the universal value of
$\rho L^2$ at the critical point depends on the aspect ratio (see
formula~\ref{eq:rhoscal}), and is listed in table~\ref{tab.1}.

\begin{figure*}
\includegraphics[clip,width=16cm]{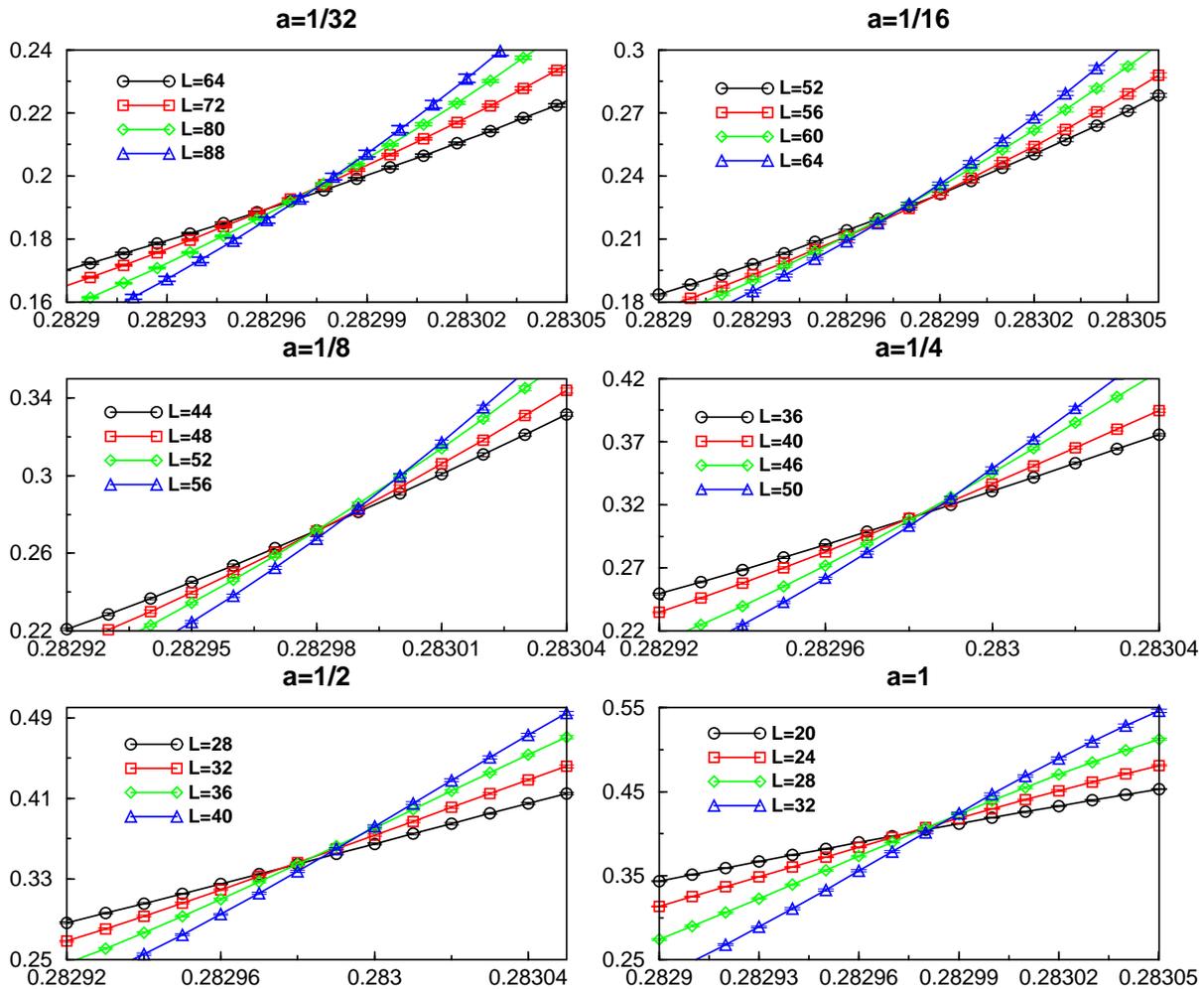}
\caption{Compressibility $\kappa$ (vertical axis) versus coupling constant $K$
  (horizontal axis) for different
  values of aspect ratio $a$ and system sizes $L$. Lines are guides to the
  eye. For all
  aspect ratios, a well-defined crossing of $\kappa$ for different system sizes is apparent, giving a
  value of $K_c$ agreeing with the estimate obtained from the crossing of $\rho
  L^2$ (figure~\ref{fig:RhoL2}).}
\label{fig:Kappa}
\end{figure*}

In figure~\ref{fig:Kappa}, we show our results for the compressibility
$\kappa$, as a function of the effective temperature $K$, for all aspect ratios simulated.
Also in this case do we
observe a well defined crossing of curves for different systems sizes as one would
expect from finite-size scaling theory. 
There is a slight variation of the estimated $K_c$ for the different values
of the aspect ratios (see table~\ref{tab.1}), but we
can safely estimate $K_c$ from the crossing of $\kappa$ to be $K_c=0.28298(2)$. This
estimate of $K_c$ agree within error bars with the one previously obtained from the
crossing of $\rho L^2$, even if they do not perfectly coincide. It is natural 
to expect such tiny deviations, and only in the thermodynamic limit should all estimates (for
all different aspect ratios and for all different estimators of $K_c$) 
give the same value for $K_c$. The precision at which we are able to calculate $K_c$ largely
exceeds previous studies~\cite{vanOtterlo94,vanOtterlo95}.

The fact that all our results in Fig~(\ref{fig:RhoL2}) and Fig.~(\ref{fig:Kappa}) show a single
well defined crossing for the large system sizes used lends strong support
support to a value of $z=2$, as predicted by the scaling theory~\cite{Fisher89c}. 
This value of $z$ was implicitly used in the simulations through the choice of the lattice
sizes and the well defined crossings implies that this choice was correct.
In the next section (\ref{sec:Huse}) of the paper, we will show more convincing numerical evidence for
this value of the dynamical critical exponent.

Another important conclusion can be drawn from the Fig~(\ref{fig:RhoL2}) and Fig.~(\ref{fig:Kappa}):
The fact that the estimates of $K_c$ from the scaling of $\rho L^2$ and $\kappa$
show a single well-defined $K_c$ allows us to rule out a very hypothetic scenario consisting of
two separate transitions with an intervening exotic phase. Our results are clearly
only consistent with a single well-defined transition.

We now turn to a discussion of our estimate of $\nu$ the correlation length exponent. In the inset of each of
the six panels in figure~\ref{fig:RhoL2}, is
shown, on a log-log scale,  the size dependence of $L^2 \frac{d\rho}{dK}$ calculated at the associated 
critical point $K_c$. From a power law fit (see
equation~(\ref{eq:drho})), we can estimate the correlation length critical
exponent $\nu$ for all aspect ratios. 
The resulting estimates are also listed in
table~\ref{tab.1}. In the present study we are able to simulate significantly larger
systems than was previously possible in particular for small values of $a$.
For $a=1/32$, we
obtain a value $\nu=0.52(2)$. This estimate is in very good agreement with the mean
field value $\nu=1/2$ given by the scaling theory of Fisher {\it et al.}'s theory~\cite{Fisher89c}. 
The small deviation observed for larger aspect ratios could be due to small finite
size deviations as we shall discuss in the subsequent section.

To conclude this section, we would like to stress once more the importance
of simulating larger systems to get more precise estimates of the critical properties.
A very slow approach to the thermodynamic limit is seen for all the data in Fig.~(\ref{fig:RhoL2}) and (\ref{fig:Kappa}).
For reasons of clarity we have not included the data for smaller sizes in these two figures. However, the smallest lattice
sizes show a clear deviation from scaling and the lattice needed to obtain a well defined crossing of the curves is in
most cases considerable.
In section~\ref{sec:aspect}, we will show that significant systematic
finite size effects could easily give rise to an incorrect interpretation of data for
certain aspect ratios.

\begin{table*}
\begin{tabular}{|c|c|c|c|c|c|c|}
\hline
Aspect ratio $a$ & $1/32$ & $1/16$ & $1/8$ & $1/4$ & $1/2$ & 1 \\ \hline
Maximum system size $L_{\rm max}$ & $88$ & $64$ & $56$ & $50$ & $40$ &
$32$ \\ \hline
$K_c$ (estimated from $\rho L^2)$ & 0.282982(5) & 0.282993(7) &
0.283000(6) & 0.282997(7) & 0.282999(6) & 0.282998(7) \\ \hline
$\rho L^2$ at $K_c$ & 0.397(17) & 0.863(70) & 1.177(60) & 1.177(60) &
0.867(40) & 0.625(25) \\ \hline
$\nu$ & 0.52(2) & 0.46(2) & 0.47(2) & 0.47(2) & 0.47(15) & 0.47(15) \\
\hline
$K_c$ (estimated from $\kappa$) & 0.282975(8) & 0.282980(12) & 0.28300(12)
&
0.282991(8)& 0.282996(8) & 0.292987(8) \\ \hline
$\kappa$ at $K_c$ & 0.196(6) & 0.227(9) & 0.299(16) & 0.328(10) &
0.374(16) & 0.415(15)\\ \hline
\end{tabular}
\label{tab.1}
\caption{Estimates of several variables for different aspect ratios
used in the simulation. See text for details.}
\end{table*}

\subsection{Scaling plot for $z$}
\label{sec:Huse}

In this section, we will show further numerical evidence for a value of $z=2$
for the dynamical exponent. In the context of quantum phase transitions in
quantum spin glasses, Huse and coworkers~\cite{Guo} (see also
Ref.~\cite{Rieger}) used a numerical method
exploiting the anisotropy in the imaginary time direction to obtain an
estimate for the dynamical exponent and the critical point. They proposed to
plot the Binder cumulant versus the aspect ratio $a=L_\tau/L^z$ for
different lattice sizes $L$ (i.e. they simulated systems of size
$L$x$L$x$L_\tau$ for different $L_\tau$). On the basis of scaling arguments
for the Binder cumulant~\cite{Guo}, it is then possible to show that all data should
collapse onto a single curve  at the critical point if the correct value of $z$ is used
to calculate $a$ from the $L_\tau$ used in the simulation. Alternatively, if one
for several different values of $L$ can locate a maximal value of the
Binder cumulant as a function of $L_\tau$, then the $L_\tau^{max}$ associated with
this maximum should scale with $L$ as $L_\tau^{max}\sim L^z$. For this to work
a very precise estimate of the critical point is presumably not necessary.

We adapt this method to the quantum phase transition studied here but use
the quantity $\rho L^z$ instead of the Binder cumulant. This quantity
displays the same properties as the Binder cumulant for the application of
the Huse method but suffers from the drawback that it depends on an initial
assumption of $z$. Please note that, the {\it a priori} unknown value of
$z$ enters implicitly in {\it both} axes in this plot. Since we have a good
estimate for that $z=2$, we first calculated $\rho L^2$ for different $L$ and $L_\tau$,
instead of trying less probable values for the dynamical exponent.

\begin{figure}
\includegraphics[clip,width=8cm]{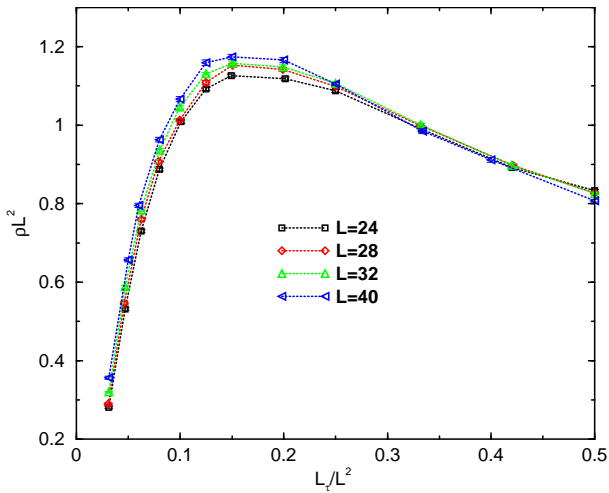}
\caption{Scaling plot of $\rho L^2$ versus aspect ratio $a=\frac{L_\tau}{L^2}$
for different system sizes at $K=0.28299$.}
\label{fig:Huse.Rho}
\end{figure}

We show the results of our calculations in figure~\ref{fig:Huse.Rho}
for a value of the effective temperature equal to
our previous estimate of the critical point $K_c=0.28299(2)$. Please note that
due to computer time restrictions, we only simulated systems of
moderate size ($L\leq 40$) for aspect ratios $a\leq 1/2$. 
Clearly all curves start to collapse into a single one, giving strong evidence
that our previous estimate for $z=2$ was correct. 
This also confirms the validity of our
previous estimate of $K_c$.
Some deviation from a perfect collapse behavior is however present and is attributed to small finite size effects since for this calculation we
only considered systems with size $L\leq 40$. In particular, small differences
in the finite-size estimate of $K_c$ (as was shown in table \ref{tab.1}) for
different aspect ratios are most probably at the origin of this small deviation. We further discuss finite size effects in the following section.

\subsection{Effects Depending on the Aspect Ratio}
\label{sec:aspect}

\subsubsection{Multiple Peaks}

\begin{figure*}
\includegraphics[clip,width=12cm]{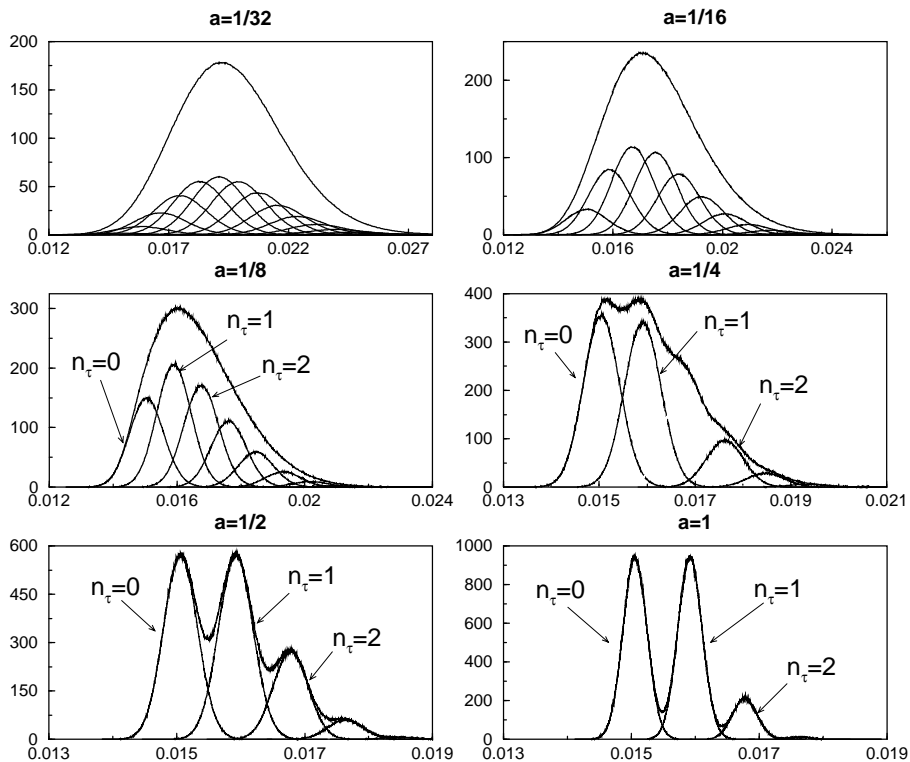}
\caption{
  Probability, $P(e)$, (vertical axis) of energy per site $e$ versus $e$
  (horizontal axis) observed during the Monte
  Carlo simulations for different aspect ratios at $K=K_c$. Here $L=32$ for all aspect
  ratios. 
  Also shown is  the
  contribution of each particle sector (see text) to $P(e)$. For the four
  larger aspect ratios, contributions of the $n_\tau=0$, $n_\tau=1$ and
  $n_\tau=2$ sectors are singled out. For reasons of clarity, we did not denote
  contributions of higher number of particle sectors. For the two smaller
  aspect ratios, the contributions of all particle sectors are also not 
  denoted for clarity reasons. }
\label{fig:Histo2}
\end{figure*}

From the results presented in the previous two sections it is clear that
the approach to the thermodynamic limit is quite slow and pronounced finite-size
effects are present at small lattice sizes. In the presence of long-range interactions
it is known that the transition in some cases can be first-order~\cite{supersolid}. 
In light of this result we verified whether the present transition also showed
signatures of a first-order transition even though
only the on-site repulsion term $U$ is included.
We study this by examining in detail the energy
distribution close to $K_c$.
In all cases we take $z=2$.
We examined carefully the
energy distribution near the critical point for the six different values
of the aspect ratio $a=1,1/2,1/4,1/8,1/16,1/32$ used in our simulations. 
Our results are shown in Fig.~\ref{fig:Histo2} where we show
the probability $P(e)$ of
observing an energy per site $e$ versus
$e$ at the critical point $K=K_c$. As is evident from
the results in Fig.~\ref{fig:Histo2} multiple peaks are present in the
energy distribution for the three largest aspect ratios while the
distribution of the energy for the three smaller aspect ratios show a single peak at the
critical point.

Given this observation, it is instructive (see Fig.~\ref{fig:Histo2}) 
to extract the contribution to $P(e)$ from the different sectors of
winding number $n_\tau$ in the geometrical worm algorithm, which can be identified with the particle number
(the number of bosons)
in the system. From the results shown in Fig.~\ref{fig:Histo2} it is clear
that for the three smallest aspect ratios, many sectors of
the particle number contribute to $P(e)$ at $K=K_c$, in particular so for the smallest
aspect ratio $a=1/32$. However for the larger aspect ratios, $a=1/4,1/2,1$,
only the sectors $n_\tau=0,1$ and eventually $2$
contribute significantly to  $P(e)$ at $K=K_c$. In particular, the two main peaks
observed in the total histogram $P(e)$ can be clearly attributed to the
contributions from the sectors with $0$ or $1$ particle in the system.
Moreover, these two peaks are clearly separated.

The generic transition is clearly driven by density fluctuations as stressed
by Fisher {\it et al.}~\cite{Fisher89c}.
If this transition is second order, many particle sectors should contribute
to $P(e)$ at the critical point. This is clearly the case for $a=1/32,1/16$ and to 
a certain extent for $a=1/8$. However, the multi-peak structure in $P(e)$ observed
for $a=1/4,1/2,1$ could be interpreted as a signature of a {\it
first order} phase transition. In particular, the energy gap between the 
contribution from the two main particle
sectors (which is the energy difference between the maxima of the two main peaks)
should remain non-zero in the thermodynamic limit for a first order transition. 
Although it would seem only a remote possibility that the order of the phase
transition could change with the aspect ratio, a more thorough analysis of the
signatures of a first-order transition observed for $a=1/4,1/2,1$ is clearly
needed, as done in the next section.

\subsubsection{Scaling of Peaks}

It is not possible to draw any definitive conclusion about the order
of the phase-transition from the energy distribution $P(e)$ calculated
for a single $L$ for the three aspect ratios $a=1/4,1/2,1$.
It is necessary to verify that this is not simply a finite size effect that
will disappear as the thermodynamic limit is approached.
Indeed, since we are doing simulations on a
finite lattice, the {\it density} of bosons will not fluctuate continuously
but only to vary it in steps of $1/L^2$. There have been many
situations where one observes evidences for a first order transition for
small lattice sizes which turns out to be second order when simulating larger
systems (see for example Ref.\cite{fuku}).

We therefore made simulations for several lattices sizes for the
aspect ratios $a=1/4,1/2$ and $1$. We observe for all cases a multi-peak
structure in the probability distribution $P(e)$. To treat correctly the
size dependence of these two peaks, one usually employs the method proposed by
Lee and Kosterlitz~\cite{Lee90} to distinguish between a first or a second
order phase transition. First, with the help of reweighting
techniques~\cite{Ferrenberg88}, we locate the temperature $K$ where the two
peaks are almost of the same height~\cite{notepics}. Then we calculate at
this temperature the free energy-like quantity $-\ln P(e)$.

\begin{figure*}
\includegraphics[clip,width=12cm]{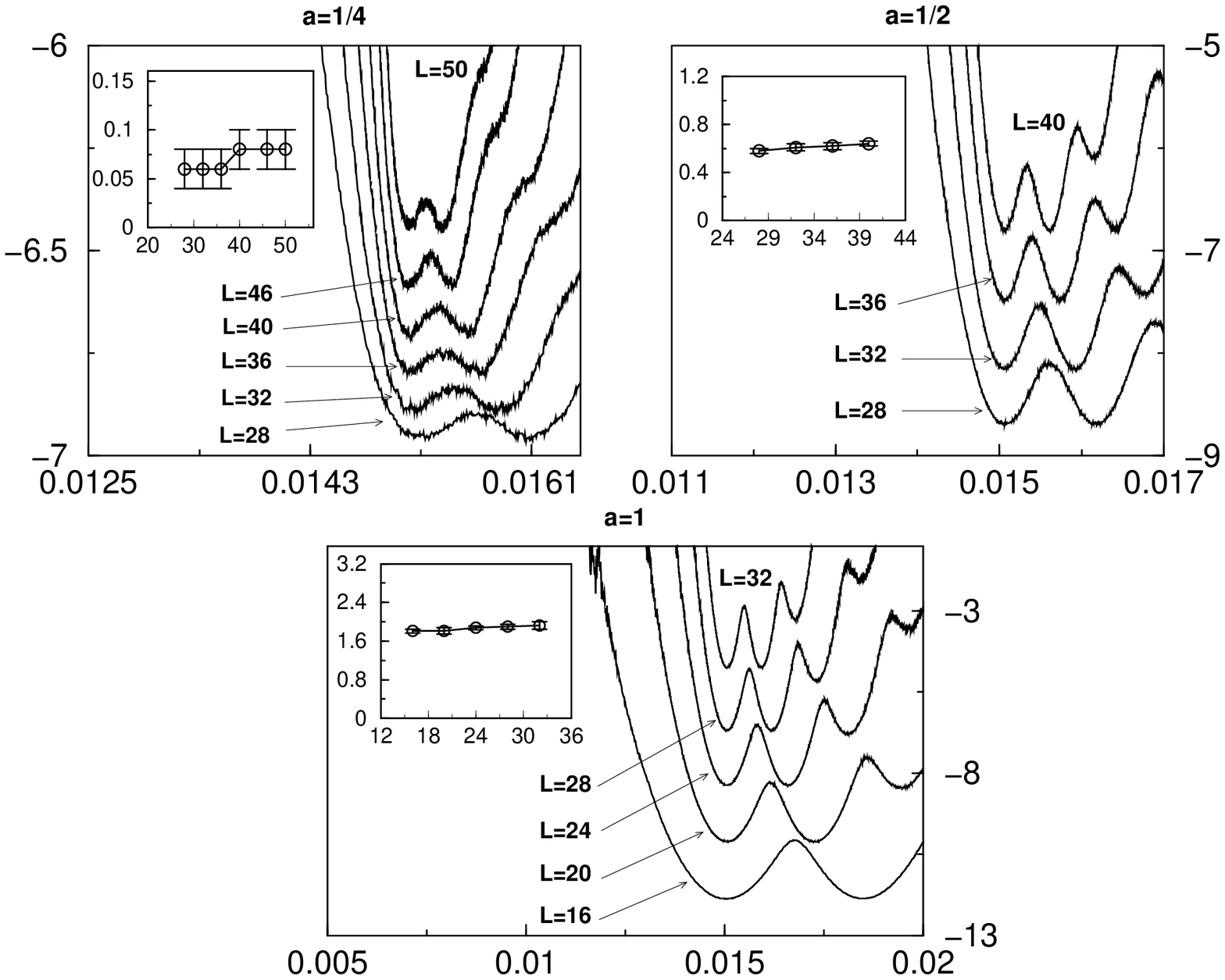}
\caption{The free energy-like quantity $-\ln{P(e)}$ (see text) for the three aspect ratios
  $a=1/4,1/2,1$ versus energy density $e$. The curves for different system
  sizes $L$ have been arbitrarily
  shifted vertically for clarity reasons. Insets : The free energy difference
  $\Delta F$ (see text) versus system size $L$.}
\label{fig:pics}
\end{figure*}

Our results are shown 
in Fig.~\ref{fig:pics}. For all the lattice sizes $L$, we still have two pronounced
peaks which seem to stay of about the same size (or actually slightly increase
for $a=1/4$, this is due in this specific case to the proximity of the
$n_\tau=2$ peak, see footnote~\cite{notepics} and below) as $L$ increases. 
More quantitatively, we plotted in the inset of figure~\ref{fig:pics} the
variation with system size of the "free energy difference"~\cite{Lee90} $\Delta
F=\ln(\frac{P^{\rm max}(e)}{P^{\rm min}(e)})$, where $P^{\rm max}(e)$ is the
height of both peaks and $P^{\rm min}(e)$ the height of the minimum between
them. We see that this quantity is constant with $L$ within error bars. 
In Lee-Kosterlitz's method~\cite{Lee90}, this indicates a {\it second
order phase transition}.

\begin{figure*}
\includegraphics[clip,width=12cm]{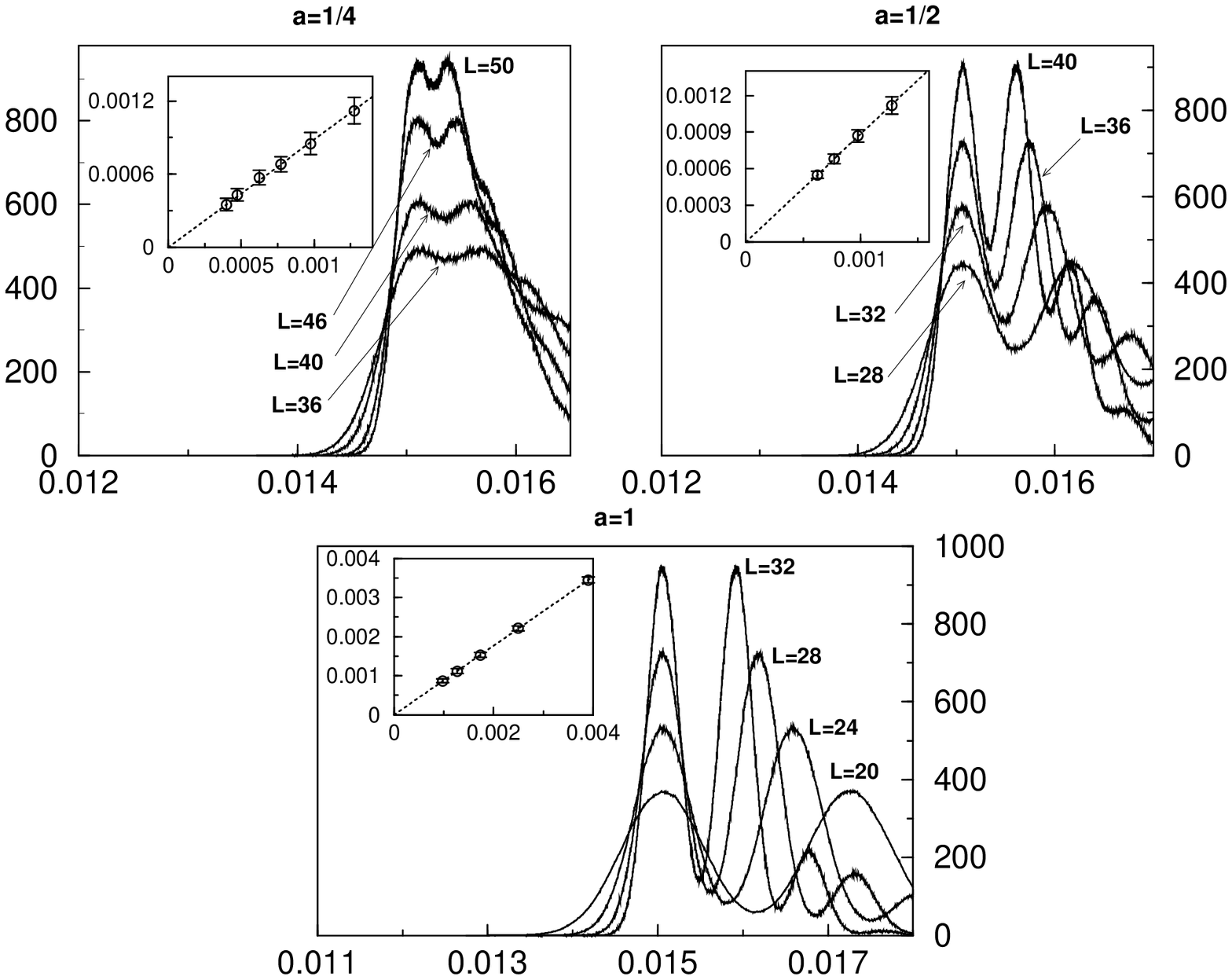}
\caption{Scaling of the peaks in $P(e)$ for different system sizes for the
  aspect ratios $a=1/4,1/2,1$. Insets : Energy gap $\Delta_{01}$ separating
the two first peaks as a function  of $1/L^2$. Dashed line is a a fit to $\Delta_{01}=A/L^2$.
}
\label{fig:pics.move}
\end{figure*}

To demonstrate this even more clearly, we also note that the separation of the peaks
(the energy gap) {\it decreases} with the system size, as can be seen in Fig.~\ref{fig:pics.move}.
In particular, we observe that whereas the first peak (corresponding to
$n_\tau =0$) stays peaked around a value $e \sim 0.015$ , the peak corresponding to
$n_\tau=1$ is clearly shifted towards the first peak as we increase system
size. Looking more carefully at the system size dependence of the energy gap
$\Delta_{01}$ between the peak for the $n_\tau =0$ and $n_\tau =1$ sectors, we
observe that it vanishes as $1/L^2$ (see insets of
Fig.~\ref{fig:pics.move}). We also observe the same scaling for the other gaps
$\Delta_{02}, \Delta_{03}$ corresponding to secondary peaks. This behavior clearly corresponds to a second
order phase transition occurring in the thermodynamic limit.  The $1/L^2$ scaling of the separation of the peaks
is presumably simply due to the fact the particle (boson) {\it density} varies as $1/L^2$ between the two peaks on the
finite lattice. In the thermodynamic limit the density can vary continuously.
In essence, the spacing between the peaks is just a reflection of the finite energy cost associated with
adding an additional boson.
Moreover, we observe that the width of the
peaks is decreasing with system size.

We conclude that the observed transition is a second order phase
transition, but with strong finite size effects for large aspect
ratios reminiscent of a first-order phase transition. These effects are presumably simply
due to the fact that the density of bosons can not be varied continuously on a finite
lattice and a result of the finite energy required to add an extra particle. 
Interestingly, these finite-size effects do not seem to be reflected
too much in the estimate of the critical point $K_c$ obtained by crossing of
$\rho L^2$ or the compressibility $\kappa$ and could easily have been missed in 
previous studies.
However, it is immediately clear that if
only the sector with particle number $n_\tau=0$ is contributing significantly at $K_c$
then we are effectively studying the transition occurring at a {\it fixed} particle number
and {\it not} the {\it generic}, incommensurate transition. The transition at fixed particle
number is equivalent to the $d+1$-$XY$ commensurate transition occuring at the tip of the MI lobes.
The non-zero chemical potential 
is decoupled and is not taken into account correctly. The critical exponents
calculated are therefore likely to be {\it strongly} influenced. 
This result is of significant importance for interpreting  what was observed in previous
simulations where only very modest lattice sizes were used~\cite{vanOtterlo94,vanOtterlo95},
and the above effects therefore likely to have been pronounced.

These finite size effects could also be of importance for previous studies of the generic
transition in the presence of weak disorder~\cite{Kisker,Park,Lee,Leerr}. The situation
is perhaps less clear here since a strong enough disorder would smear the multi-peak structure
observed above and enhance particle number fluctuations. However, weaker disorder would only
broaden the individual peaks slightly and one would effectively be studying the influence of
disorder on the {\it commensurate} transition occurring at $\mu=0$ (albeit at fixed particle number).
Recent calculations~\cite{ProkDis1,ProkDis2} done with a
variant~\cite{Prokworm} of the worm algorithm used in this work
have shown that the situation in this case is much more subtle than previously thought,
questioning the validity of older work considering the influence of weak disorder
at the commensurate transition~\cite{Kisker,Park,Lee,Leerr}. In fact, it was observed
that the universal behavior only sets in at {\it very} large lattice sizes.
Hence, if only  weak disorder is considered for small lattice sizes one is therefore likely to
observe incorrect exponents. This would explain why Park {\it et al.}~\cite{Park} observe $\nu=0.5\pm 0.1$
at the generic transition in the presence of weak disorder violating the inequality $\nu\ge 2/d$ and
would question the validity of the phase-diagram presented by Lee {\it et al.}~\cite{Lee,Leerr}.

\subsection{Correlation Functions}
Even though in the previous two sections we have seen that the pronounced finite-size effects
present at the larger lattice sizes have relatively little influence on $\kappa$ and $\rho L^2$
they are quite visible in the correlation functions. The correlation functions can be easily calculated
using the geometrical worm algorithm as described in reference~\onlinecite{directed}. From scaling
theory~\cite{Fisher89c,Wallin} we expect them to follow the following form at $K=K_c$:
\begin{eqnarray}
C_x(x)\sim x^{-(d+z-2+\eta)}\nonumber\\
C_\tau(\tau)\sim \tau^{-(d+z-2+\eta)/z}.
\end{eqnarray}

\begin{figure}
\includegraphics[clip,width=8cm]{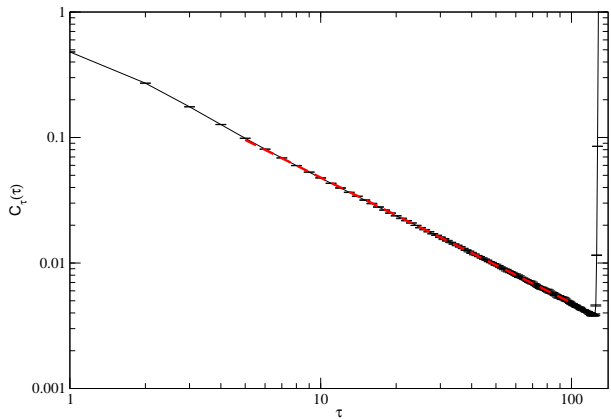}
\caption{The $\tau$-dependent correlation function calculated on a
lattice with $L=64$ and $L_\tau=128$ at $K_c$.
The dashed line (hidden by the data points) indicates a fit to the form $C_\tau(\tau)=0.47742 \tau^{-0.998}$.}
\label{fig:ct}
\end{figure}

In Fig.~\ref{fig:ct} we show representative results for $C_\tau(\tau)$ calculated at $K=K_c$
for the generic transition at $\mu=1/4$. 
There is a clear power-law dependence and it is relatively easy to extract the associated exponent
$C_\tau(\tau)=0.47742 \tau^{-0.998}$. If we take our previous estimate of $z=2$ for granted and remember
that $d=2$ this would imply that $\eta=0$ consistent with mean-field behavior.

\begin{figure}
\includegraphics[clip,width=8cm]{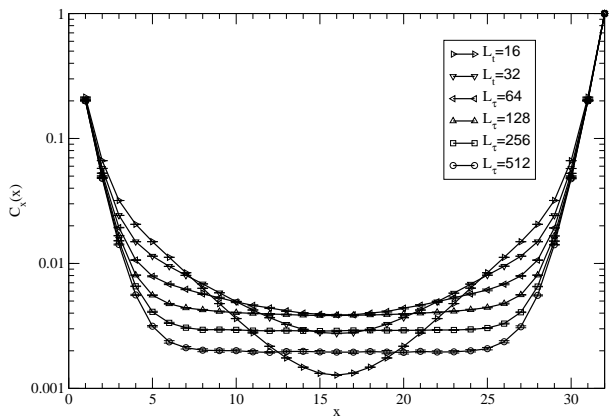}
\caption{The x-dependent correlation function at $K=K_c$ for $L=32$ calculated for
a range of different $L_\tau=16,\ldots,512$.}
\label{fig:cx}
\end{figure}

The behavior of the correlation functions in the spatial direction is much more complicated. In Fig.~(\ref{fig:cx})
we show representative results for $C_x$ calculated at $K=K_c$ for a system of size $L=32$ for a range
of different $L_\tau=16,32,64,128,256,512$.
As is clearly evident from this figure these correlation functions do not follow a power-law behavior.
Given the results presented in the previous sections this is not surprising and illustrates the difficulties
associated with these finite-size effects. The deviation from power-law behavior increases dramatically
with the aspect ratio. 
Hence, only by studying significantly larger system sizes at much smaller
aspect ratios would one presumably be able to recover the expected power-law behavior. We have so far been
unable to do so.

\section{Conclusion~\label{sec:conclusion}}
In conclusion, we have in the present paper presented large-scale numerical results for the SF-I transition occurring
in the boson Hubbard model at an incommensurate chemical potential $\mu=1/4$ obtained using the ``directed"
version of the recently developed geometrical worm algorithm~\cite{bosons,directed}. By carefully analyzing
the probability distribution $P(e)$ of the energy density at $K=K_c$ for different values of the aspect
ratio, we showed that strong finite-size effects, reminiscent of a first order transition are present for
the larger aspect ratios. These effects disappear in the thermodynamic limit and the transition is indeed
second order with mean field exponents as predicted by scaling theory~\cite{Fisher89c}. However, if only
small lattice sizes are used then these effects are pronounced and would imply that the chemical potential
is not taken into account correctly, resulting in the associated critical exponents being calculated incorrectly. 
Finally, we note that amplitude fluctuations, absent from the Hamiltonian Eq.~(\ref{eq:hV}) used here, could be
important at the generic transition as recent studies indicate~\cite{Chamon}. 

\acknowledgments
This research is supported by NSERC of Canada as well as SHARCNET. 
FA acknowledges support from the Swiss National Science Foundation.

\bibliography{muquarter}

\begin{thebibliography}{50}
\bibitem[{not()}]{notepics}
\bibinfo{note}{For the case of $a=1/4$, the value of $K$ for which the two main
  peaks of the {\it total} energy distribution $P(e)$ are of the same height is
  not strictly equal to the one for which both peaks for $n_\tau=0$ and
  $n_\tau=1$ sectors are of the same height. This is due to the viscinity of
  the $n_\tau=2$ peak that contributes to the second peak in $P(e)$ (this can
  be clearly be seen in figure~\ref{fig:Histo2}). This is not the case for
  $a=1/2$ and $a=1$ for which the $n_\tau=2$ peak is further away for the
  systems sizes studied.}
\end{thebibliography}

\end{document}